\documentclass[prd,floatfix,nofootinbib,superscriptaddress,twocolumn,showpacs]{revtex4}

\usepackage{varioref,xr} %

\usepackage{amssymb,amsfonts,amsmath,paralist} %
\usepackage{graphicx,color}

\newcommand{\CH}{\mathcal{H}} \newcommand{\dm}{\Delta \mu} %
\newcommand{\tr}{{\mathrm{tr}}} %
\newcommand{\eff}{{*}} 

\renewcommand{\max}{{\text{max}}} 
\newcommand{\mev}{\:\mathrm{MeV}} 
\newcommand{\gev}{\:\mathrm{GeV}} 
\newcommand{\const}{\mathrm{const}} %
\newcommand{\parfrac}[2]{\left(\frac{#1}{#2}\right)}
\newcommand{\pfrac}[2]{\frac{\partial #1}{\partial #2}}

\IfFileExists{srcltx.sty}{\usepackage[active]{srcltx}}

\begin{document}
\title{Self-consistent Evolution of Magnetic Fields and Chiral Asymmetry in
  the Early Universe}

\author{Alexey~Boyarsky}%

\affiliation{Instituut-Lorentz for Theoretical Physics, Universiteit Leiden,
  Niels Bohrweg 2, Leiden, The Netherlands} 

\affiliation{Ecole Polytechnique F\'ed\'erale de Lausanne, FSB/ITP/LPPC, BSP
  720, CH-1015, Lausanne, Switzerland}

\affiliation{Bogolyubov Institute of Theoretical Physics, Kyiv, Ukraine} %

\author{J\"{u}rg Fr\"{o}hlich} %
\affiliation {Institute of Theoretical Physics, ETH H\"{o}nggerberg, CH-8093
  Zurich, Switzerland}

\author{Oleg Ruchayskiy}%
\affiliation{CERN Physics Department, Theory Division, CH-1211 Geneva 23,
  Switzerland}

\begin{abstract}
  We show that the evolution of magnetic fields in a primordial plasma, filled
  with Standard Model particles, at temperatures $T \gtrsim 10\mev$ is
  strongly affected by the quantum chiral anomaly -- an effect that has been
  neglected previously. Although reactions equilibrating left and right-chiral
  electrons are in deep thermal equilibrium for $T\lesssim 80$~TeV, an
  asymmetry between these particle develops in the presence of strong magnetic
  fields. This results in magnetic helicity transfer from shorter to longer
  scales.  This also leads to an effective generation of lepton asymmetry that
  may survive in the plasma down to temperatures $T\sim 10$~MeV, which may
  strongly affect many processes in the early Universe.  Although we report
  our results for the Standard Model, they are likely to play an important
  role also in its extensions.
\end{abstract}
 

\maketitle

 
Magnetic fields are expected to play an important role in the early Universe
Recent observational indications of the presence of magnetic fields in the
inter-galactic medium~\cite{Neronov:10a,Tavecchio:10,Dolag:10} suggest that
cosmological magnetic fields (CMF) may survive even till the present
epoch. Thus they could have played the role of seeds for the formation of
galactic magnetic fields.  A number of mechanisms for the creation of CMF at
very high temperatures have been proposed (see
e.g.~\cite{Grasso:2000wj,Giovannini:2003yn,Kandus:10} and refs. therein).

In this paper we concentrate, however, on a different problem: we \emph{assume} that
strong CMF were \emph{already} generated at a temperature $\gtrsim 100\gev$ and we study the
subsequent evolution of such fields. Usually, this evolution is described by
the system of Maxwell plus Navier-Stokes equations~(for a detailed review
see~\cite{Giovannini:2003yn,Biskamp:97}).  Here we will argue that, for
temperatures $T \gtrsim 10 \mev$, this system of MHD equations \emph{should be
  extended to include a new effective degree of freedom}, even if all
particles and reactions are described by just the Standard Model of particle
physics.  This significantly affects the evolution of CMF and the state of the
primordial plasma.

At such temperature rates, of all perturbative processes related to the
electron's finite mass are suppressed as $(m_e/T)^2$. Ignoring these
corrections for a moment, the number of left and right-chiral
electrons\footnote{More precisely, $n_L$ is the \emph{difference} between the
  number of left particles and left anti-particles (same for $n_R$).}  is
conserved independently.\footnote{The number of left-chiral electrons is not
  conserved when weak processes are fast (the conserved quantities are $n_L +
  n_{\nu_e}$ and $n_R$). The coefficient $c_\Delta$ in Eq.~(\ref{eq:4}) below
  takes this into account.}  %
That is, apart from the vector current $j^\mu = \bar\psi \gamma^\mu \psi$
describing conservation of electric charge ($n_L + n_R$),
the average number density of the left- (right) chiral electrons $n_{L,R} =
\frac1{2V}\int d^3x\,\psi^\dagger (1\pm\gamma_5)\psi$ does not change with
time.  This is true on time scales smaller than the \emph{chirality-flipping
  scale} $\Gamma_f^{-1}$.  Although the chirality-flipping rate is suppressed
as compared to the rate of chirality-preserving weak and electromagnetic
processes, it is faster than the Hubble expansion rate, $H(T)$, for
temperatures below 80~TeV~\cite{Campbell:92} and chirality flipping processes
are in thermodynamic equilibrium.  Yet on time scales $\Gamma_{EM, weak}^{-1}
< t < \Gamma_f^{-1}$ one should introduce independent chemical potentials,
$\mu_L$ and $\mu_R$, for two approximately conserved number densities, with
$n_{L,R} = \frac{\mu_{L,R}}6 T^2$.  In the presence of external
\emph{classical} fields the conservation of the axial current is spoiled,
however, by the \emph{chiral anomaly}~\cite{Treiman:85} -- a quantum effect
leading to a change of $n_L - n_R$:
\begin{equation}
  \label{eq:1}
  \frac{d(n_L - n_R)}{dt} =
  \frac{2\alpha}{\pi}\frac1V\int d^3 x\, E\cdot B =
  -\frac{\alpha}{\pi} \frac{d\CH}{dt}\;,
\end{equation}
where $\alpha = \frac{e^2}{4\pi}$ is the fine-structure constant and $\CH$ is
the \emph{magnetic helicity} defined as
\begin{equation}
  \CH(t) = \frac1V\int_V d^3x \, A\cdot B\;,
  \label{eq:2}
\end{equation}
(where $B$ is the magnetic field and $A$ the vector potential, with $B =
\nabla \times A$).  The quantity~(\ref{eq:2}) is gauge invariant, provided
that $B$ is parallel to the boundary of $V$ (see e.g.~\cite{Biskamp:97}).  The
time evolution of $\CH(t)$ is given by~\cite{Biskamp:97}
\begin{equation}
  \label{eq:3}
  \frac{d\CH}{dt} = -\frac2V\int_V d^3x\,  E\cdot B\;.
\end{equation}
In terms of the difference of left and right chemical potentials, $\dm\equiv
\mu_L - \mu_R$, Eq.~(\ref{eq:1}) reads
\begin{equation}
  \label{eq:4}
  \frac{d(\dm)}{dt} = - \frac{c_\Delta \alpha}{T^2} \frac{d \CH(t)}{dt}\;,
\end{equation}
where $c_\Delta$ is a numerical coefficient of order one that describes the
dependence of $n_L$ on globally conserved charges in the primordial plasma.

If $\Delta\mu\neq 0$, the chiral anomaly leads to an additional contribution
to the current in Maxwell's
equations~\cite{Vilenkin:80a,Joyce:97,Redlich:1984md,Giovannini:1997eg,Alekseev:98a,Frohlich:2000en,Frohlich:2002fg,Fukushima:08}:
\begin{displaymath}
    \nabla\times B = \sigma E + \frac\alpha{\pi}\Delta\mu(t) B\;,
\end{displaymath}
or, combining it with the Bianchi identity $\nabla \times E = -\dot B$:
\begin{equation}
  \label{eq:5}
  \frac{\partial B}{\partial t} =   \frac1{\sigma}\nabla^2 B +
  \frac{\alpha}\pi\frac{\Delta\mu}{\sigma}   \nabla\times B\;.
\end{equation}
As weak reactions are fast enough at these temperatures to establish local
thermodynamic equilibrium (LTE), in the background of long-wavelength
electromagnetic fields, space-dependent chemical potentials $\mu_{L,R}(x)$ may
be defined. Eqs.~(\ref{eq:1}), (\ref{eq:4}) can then be written in a local
form, and Eq.~(\ref{eq:5}) acquires additional terms, proportional to the
gradients of
$\dm(x)$~\cite{Giovannini:1997eg,Frohlich:2000en,Frohlich:2002fg}. We assume
fields to be slowly varying and neglect these effects as well as those
depending on the velocity field. We will show that, even in this limit, the
evolution of magnetic fields significantly changes as compared to the usual
Maxwell equations. A more realistic analysis should include all the derivative
terms, as well as the Navier-Stokes equation describing, in particular,
turbulent effects known to be important for the evolution of CMF. We leave a
more complete microscopic derivation and an analysis of the full system to
future work and use the simple model described above to illustrate the
previously neglected effects.

Eqs.~(\ref{eq:4}--\ref{eq:5}) remain valid in an expanding Universe if written
in conformal coordinates (see
e.g.~\cite{Joyce:97,Banerjee:04,Giovannini:2003yn}). Henceforth we use
conformal quantities and define conformal time as $\eta = \frac{M_*}T$, where
$M_* = \sqrt{\frac{90}{8\pi^3 g_\eff}}M_{Pl}$ and $g_\eff$ is the effective
number of relativistic degrees of freedom.

Eqs.~(\ref{eq:4}--\ref{eq:5}) are translation and rotation invariant. We
introduce the \emph{magnetic helicity density}, $\CH_k$, and the
\emph{magnetic energy density}, $\rho_k$, in Fourier-space, with $\rho_B(\eta)
= \int dk\,\rho_k(\eta)$ and $\CH(\eta) = \int dk\,\CH_k(\eta)$.\footnote{In
  terms of left and right circular polarized modes $B_k^\pm$, $\CH_k =
  \frac{k}{2\pi^2} (|B_k^+|^2 - |B_k^-|^2)$ and $\rho_k = \frac{k^2}{(2\pi)^2}
  (|B_k^+|^2 +|B_k^-|^2)$. Integrals over $k$ run over the radial direction
  only (see e.g.~\cite{Banerjee:03,Banerjee:04,Campanelli:07}).}  The quantities $\CH_k$
and $\rho_k$ obey the inequality $|\CH_k| \le \frac{2}k \rho_k$, which is
saturated for field configurations known as \emph{maximally helical fields}.
In our subsequent analysis, we focus on this case and \emph{choose for
  definiteness} $\CH_k > 0$ and $\Delta\mu>0$.  Multiplying the Fourier
version of Eq.~(\ref{eq:5}) by the complex-conjugate mode $\vec B_k^*$, we
obtain, after some simple manipulations (see
Appendix~\ref{sec:derivation-eq-Hk} for details, cf.~\cite{Campanelli:07}),
\begin{align}
  \label{eq:6}
  \pfrac{\CH_k}{\eta} &= -\frac{2k^2}{\sigma_c}\CH_k +
  \frac\alpha\pi\frac{k\Delta\mu}{\sigma_c} \CH_k\;,\\
  \label{eq:7}
  \frac{d(\Delta\mu)}{d\eta} &= -(c_\Delta\alpha)\int dk\,\pfrac{\CH_k}\eta
  -\Gamma_f \dm \;,
\end{align}
where we have restored the chirality flipping rate $\Gamma_f$ in
Eq.~(\ref{eq:7}) and  used the conductivity $\sigma_c \equiv
\sigma(\eta)/T\approx 70$~\cite{Baym:1997gq}.

The system~(\ref{eq:6}--\ref{eq:7}) has been previously studied in two
regimes. It was demonstrated in~\cite{Joyce:97,Frohlich:2002fg,Semikoz:04a}
that, in the presence of a large initial chemical potential difference
$\Delta\mu(\eta)>0$, the quantity
\begin{equation}
  \label{eq:8}
  \CH_k(\eta) = \CH_k^0\exp\left\{\frac{2k}{\sigma_c} \Bigl(
    \frac{\alpha }{2\pi}\int^\eta_{\eta_0} \hskip -1ex\Delta\mu(\tilde \eta)d\tilde\eta -
    k(\eta-\eta_0) \Bigr)
  \right\}
\end{equation}
\emph{grows exponentially fast for sufficiently long wavelengths}. Conversely,
in~\cite{Giovannini:1997eg} the initial background of helical (hyper)magnetic
fields was used to generate a non-zero chemical potential for $T>100\gev$.

In this work, however, we consider helical CMF with some initial spectrum
$\CH_k^0$, already present at $T\sim 100\gev$ in the hot plasma, filled with
particles in thermal equilibrium
(cf.~\cite{Banerjee:03,Banerjee:04,Campanelli:07,Jedamzik:10}). It was believed that as
$\Gamma_f \gg H(T)$ for $T\lesssim 80$~TeV, no chiral asymmetry will survive.

\bigskip

\textbf{Chirality evolution.} \emph{Below we show} that both $\dm$ and the
magnetic helicity do survive below $100\gev$ on time-scales much longer than
diffusion or chirality flipping times (till $10\div 100\mev$). (For $\Gamma_f
\to 0$ the system~(\ref{eq:6}--\ref{eq:7}) can even reach a stationary state
with non-zero $B$ and $\dm$).

To see this, it is convenient to separate on the right hand side of
Eq.~(\ref{eq:7}) a source term $S_B(\eta)$ (independent of
$\dm$) (cf. \cite{Giovannini:1997eg}):
\begin{equation}
  \label{eq:9}
  \frac{d(\Delta\mu)}{d\eta} = -(\Gamma_B(\eta)+\Gamma_f)\Delta\mu + S_B(\eta)\;,
\end{equation}
where
\begin{equation}
  \label{eq:10}
  \begin{aligned}
    \Gamma_B(\eta) &\equiv \frac{c_\Delta\alpha^2}{\pi \sigma_c} \int
    dk\,k\CH_k = \frac{2c_\Delta\alpha^2}{\pi \sigma_c}\rho_B \;,\\
    S_B(\eta) &\equiv 2\frac{c_\Delta\alpha}{ \sigma_c} \int dk\,k^2 \CH_k\;.
  \end{aligned}
\end{equation}

We begin our analysis of Eqs.~(\ref{eq:6}) and (\ref{eq:9}--\ref{eq:10}) with
the case where $\Gamma_f = 0$ and the field is \emph{initially
  ``monochromatic''}, i.e.,
\begin{equation}
  \CH_k(\eta) = \CH(\eta)\delta(k-k_0)\;.
\label{eq:11}
\end{equation}
 The form~(\ref{eq:11}) is preserved during the
evolution as Eq.~(\ref{eq:6}) is homogeneous.\footnote{\label{fn:1}This  is an
  artifact of our homogeneous approximation~(\ref{eq:4}--\ref{eq:5}), with
  $\dm$ independent on spatial coordinates, see below.}
Putting
in Eq.~(\ref{eq:9}) $d(\dm)/d\eta = 0$ and  $\Gamma_f=0$ we find the
so-called \emph{tracking solution}:
\begin{equation}
  \label{eq:12}
  \Delta\mu_{\tr} = \frac{S_B(\eta)}{\Gamma_B(\eta)} = \frac{2\pi k_0}\alpha \;.
\end{equation}
This is an \emph{exact static} solution of the
system~(\ref{eq:6}),~(\ref{eq:9}): $\dm_\tr$ and $\CH(\eta)$ \emph{remain
  constant}, i.e., dissipation due to magnetic diffusion is \emph{exactly
  compensated} by growth due to a non-vanishing chemical potential difference
$\dm_\tr$; (cf.~(\ref{eq:8})).

Until now we have completely neglected the massiveness of the electrons. 
It is straightforward to compute that the rate $\Gamma_f(\eta)$ due to
electromagnetic processes is:\footnote{The contribution of weak processes to
  $\Gamma_f$ is small at $T < 100\gev$, see
  Appendix~\ref{sec:pert-chir-flipp}.}  $ \Gamma_f(\eta) \approx
\alpha^2 \parfrac{m_e}{3M_*}^2\eta^2 $.  Eqs.~(\ref{eq:6}), (\ref{eq:9}) can
be rewritten to describe deviations from the equilibrium static
solution~(\ref{eq:12}):
\begin{align}
  \label{eq:14}
  \frac{d \Delta \mu}{d\eta} 
  &= -\Gamma_B (\Delta \mu - \Delta\mu_\tr) - \Gamma_f \dm\;,\\
  \label{eq:15}
  \frac{d \Gamma_B}{d\eta} &=
  \frac{\Gamma_B}{\eta_\sigma}\left(\frac{\dm}{\dm_\tr} - 1\right)\;,
\end{align}
where $\eta_\sigma \equiv \frac{2k^2}{\sigma_c}$ is the \emph{magnetic
  diffusion time}.  From Eq.~(\ref{eq:14}) we see that $\Gamma_B$ and
$\Gamma_f$, which enter symmetrically in Eq.~(\ref{eq:9}), play very different
roles.  The rate $\Gamma_f$, that depends \emph{only} on temperature,
constantly drives $\dm$ to zero.  The $\Gamma_B$ term pushes the system
towards the equilibrium value~(\ref{eq:12}) (that depends only on $k_0$). It
depends on $\rho_B$ and has its own dynamics, Eq.~(\ref{eq:15}).

If the magnetic field is large (such that $\Gamma_B \gg \Gamma_f$), any initial
value of $\dm$ will be quickly ``forgotten'' and $\dm$ will be driven towards
$\dm_\tr$. At that moment a new tracking solution will take over, with $\dm - \dm_\tr \approx \gamma \dm_\tr$, where
\begin{equation}
  \label{eq:16}
  \gamma(\eta) \equiv \frac{\Gamma_f(\eta)}{\Gamma_B(\eta)} \;.
\end{equation}
This new solution is valid, provided two conditions hold: \textit{(i)}
$\gamma\ll \Gamma_B t$ and \textit{(ii)} $\gamma \ll \Gamma_B \eta_\sigma$.
When this holds the evolution of $\Gamma_B$ is given by~(\ref{eq:15}).
\begin{equation}
  \label{eq:17}
  \frac{d \Gamma_B}{d\eta} = -\frac{\gamma(\eta)}{\eta_\sigma}\Gamma_B =
  -\frac{1}{\eta_\sigma}\Gamma_f(\eta) \;.
\end{equation}
Eq.~(\ref{eq:17}) shows that $\Gamma_B$ remains practically constant when
$\eta\ll \eta_\sigma/\gamma(\eta)$, which is \emph{significantly longer} than
$\eta_\sigma$, as $\gamma \ll 1$. To estimate the time at which the function
$\gamma(\eta)\sim 1$ we note that it evolves with time because of an
increasing chirality flipping rate $\Gamma_f(\eta) \propto \eta^2$ and because
the total magnetic energy dissipates~(\ref{eq:17}).  Neglecting this latter
change, we estimate $\gamma$ to be given by:
\begin{equation}
  \label{eq:18}
  \gamma = \frac{\pi\sigma_c}{2c_\Delta }\parfrac{m_e}{3 M_*}^2\frac{\eta^2}{
    \frac{\pi^2}{30}g_\eff\;r_B } =
  \frac{10^{-5}}{r_B}\parfrac{100\mev}{T}^2\parfrac{30}{g_\eff}\;,
\end{equation}
where we used $r_B \equiv \rho_B/(\frac{\pi^2}{30}g_\eff T^4)$ is the fraction
of magnetic energy density to the total energy density.  From
Eq.~(\ref{eq:17}) we see in addition that $\Gamma_B$ remains approximately
constant as long as $ \frac1{\eta_\sigma}\int \gamma(\eta)d\eta < 1 $.
Using~(\ref{eq:18}) we find that (this is illustrated in
Fig.~\ref{fig:helicity_flip_gamma1} in Appendix~\ref{sec:evol-single-mode}).
\begin{equation}
 \label{eq:19}
  \frac{\gamma(\eta) \eta}{3 \eta_\sigma} \le 1\;.
\end{equation}
%

\begin{figure*}[t]
  \centering
  \begin{tabular}{cc}
    \includegraphics[width=.5\textwidth]{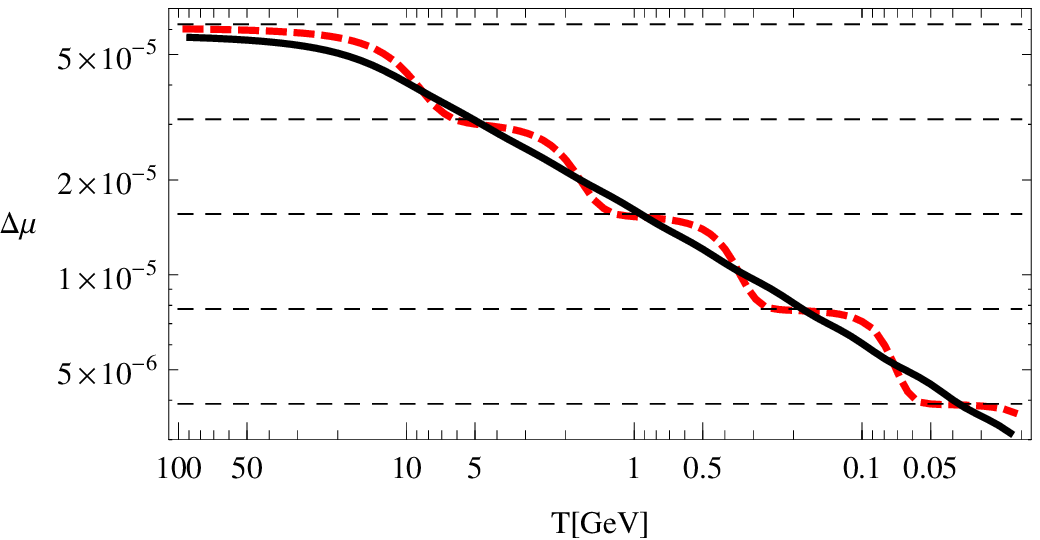}    &
    \includegraphics[width=.5\textwidth]{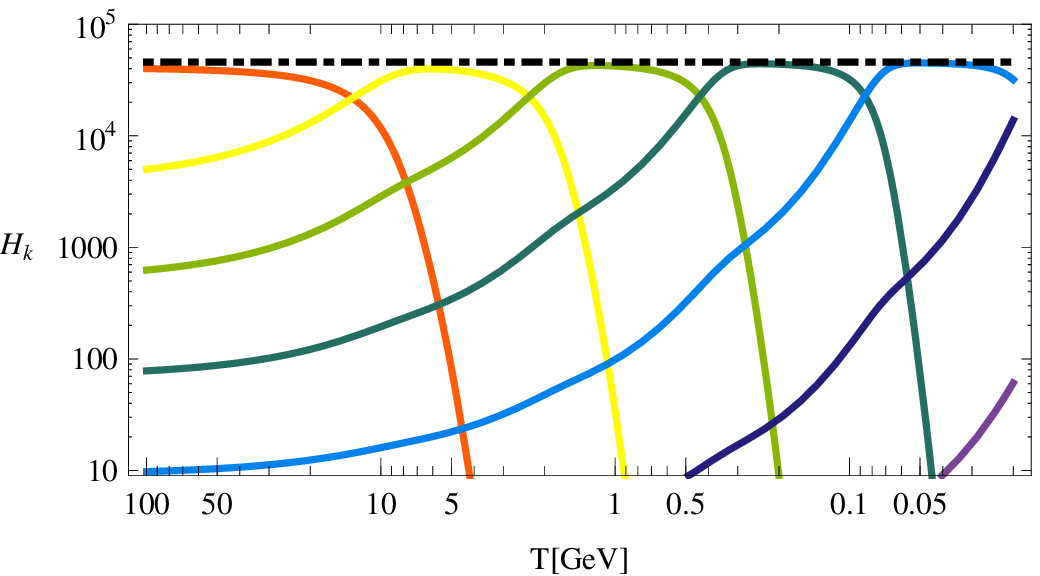}
  \end{tabular}
  \caption{Evolution in the absence of chirality flip for continuous spectrum
    $\CH_k^0 \propto k^3$.  \textbf{Left:} the evolution of $\dm(\eta)$ (black
    solid line). The red dashed line: the approximation of the spectrum by 10
    modes ($k_n = 10^{-10}\times 2^{n-1/2}$, $n=1,10$).  The horizontal lines
    are tracking solutions for individual modes $k_n$.  \textbf{Right:}
    transfer of helicity from the shorter to the longer modes (red to blue).}
  \label{fig:helicity}
\end{figure*}

\bigskip

\noindent\textbf{Inverse cascade.} 
So far we have considered a toy model example of "monochromatic" helical field
(\ref{eq:11}). Although the Eq.~(\ref{eq:6}) is linear, the modes $\CH_k$ are
not independent for different $k$ (due to the integral in the
Eq.~(\ref{eq:7})).  For a continuos spectrum, this interaction results in
another very important effect: the initial spectrum reddens with time, the
total helicity being conserved (similarly to the ``\emph{inverse cascade}''
phenomenon~\cite{Biskamp:97}, Sec.~7.2.3.

Indeed, let us consider first the case of two
modes $(k_1,\CH_1(\eta))$ and $(k_2, \CH_2(\eta))$ with $k_1 > k_2$, to
understand the situation qualitatively. While $\Gamma_B \gg \Gamma_f$, the
evolution for $\Delta\mu$ has the form:
\begin{equation}
  \label{eq:20}
  \frac{d(\Delta\mu)}{d\eta} = -\frac{c_\Delta\alpha^2}{\pi  \sigma_c}
  \Bigl(k_1
  \CH_1 + k_2 \CH_2\Bigr)\Delta\mu + \frac{2c_\Delta\alpha}{ \sigma_c}\Bigl(k_1^2
  \CH_1 + k_2^2 \CH_2\Bigr)\;.
\end{equation}
One can again try to construct a tracking solution of Eq.~(\ref{eq:20}) by
putting its l.h.s. to zero.  It is clear, however, that, unlike in the
case~(\ref{eq:12}), such a tracking solution cannot be time
independent. Indeed, according to Eq.~(\ref{eq:6}) $\dot \CH_k = 0$ only if
$\dm = \frac{2\pi k }\alpha$, while our solution $\dm_\tr =
\frac{2\pi}{\alpha}\frac{k_1^2 \CH_1 + k_2^2 \CH_2}{k_1 \CH_1 + k_2 \CH_2}$
depends on both modes.  In the case where a shorter mode ($k_1$) contains most
of the energy density, $\dm$ will grow very fast and reach: $\dm_\tr \approx
\frac{2\pi k_1}\alpha(1-\epsilon) $.  Initially $\epsilon = \frac{k_2
  \CH_2}{k_1\CH_1} \ll 1$ as $\CH_2$ is subdominant and $\dm$ is close to its
``static'' value for $k_1$. Therefore the mode $\CH_1$ remains almost constant,
for $\eta < \eta_\sigma(k_1)/\epsilon(\eta)$. For the mode $\CH_2$, however,
$k_2< \frac{\alpha \dm_\tr}{2\pi}$ and from the solution~(\ref{eq:8}) (valid
for any $\dm(\eta)$) we see that $\CH_2$ will start growing. As its growth
enters the exponential phase, $\epsilon$ increases and $k_1$ becomes
\emph{greater than} $\dm(\eta)$, causing $\CH_1$ to decay exponentially.
$\dm(\eta)$ will therefore quickly evolve to the value $\frac{2\pi}\alpha
k_2$.  From Eq.~(\ref{eq:6}) we find that for $\epsilon \ll 1$:
\begin{equation}
  \label{eq:21}
  \CH_2(\eta) \approx \CH_2(\eta_0)e^{\frac{2 k_1 k_2}{\sigma_c}\eta}
\end{equation}
and see that $\dot \CH_1 = - \dot \CH_2$ \emph{as long as $\Gamma_B\gg
  \Gamma_f$}, i.e. the total helicity of the system is
conserved.\footnote{$\dot\CH$ is related to $\dot\dm$ by Eq.~(\ref{eq:4}).
  Even if $\dm$ changes smoothly, the very small numerical coefficient in
  (\ref{eq:4}) suppresses the change of $\CH$.}
  
The  evolution of continuous spectra is  qualitatively very
 similar.  Assume that the initial helicity spectrum $\CH_k^0$ has its
maximum at a scale $k_1$ and then decays as $\CH_k^0 \propto
(\frac{k}{k_1})^{n_s-2}$, with $n_s \ge 3$. The scale $k_1$ determines the
value of $\dm$ at the beginning, while the longer modes grow.
At the moment when back-reaction of these growing modes on $\dm$ becomes
non-negligible, the chemical potential difference gets smaller and the modes
with $k_1 \gtrsim k > \frac{\alpha\dm (\eta)}{2\pi}$ start decaying.

For discrete spectrum $\dm$
changes by ``steps'', defined by the modes $k_n$ (see
Fig.~\ref{fig:helicity} for $\Gamma_f = 0$, red dashed curve). Every
such ``step'' corresponds, on the other panel of Fig.~\ref{fig:helicity}, to a
fast decay of one helicity mode and exponential growth of an adjacent one
(from red to purple), while the total helicity remains constant (black
dot-dashed line). 
The conservation of helicity implies that the total magnetic energy gets
dissipated as $\rho_k = \frac k2 \CH_k$ for helical fields.
If we sample the same spectrum with a larger number of
modes, the evolution of $\dm$ becomes monotone. If the initial spectrum is
sharp ($n_s > 3$), $\dm$ will decay more slowly, and the short modes will
survive for a longer time.  The resulting spectrum will, however, be roughly
the same -- the helicity concentrates around the longest mode $k_2$ that had
enough time to start growing, $\dm = \dm_\tr(k_2)$, the magnetic energy is
smaller by the factor $\frac{k_2}{k_1}$.  We believe that these results
correctly describe the interaction between different helicity modes even in
the inhomogeneous case, provided that deviations from LTE are not very
dramatic and $\dm(x)$ is smooth.

%
\begin{figure}[t]
  \centering
  \includegraphics[width=.5\textwidth]{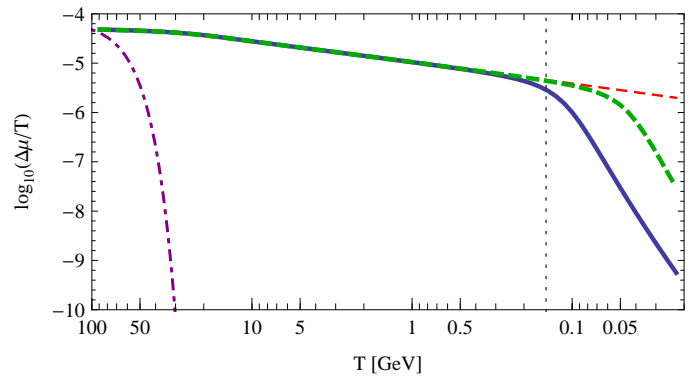}
  \caption{Evolution of the chemical potential for $r_B = 5\times 10^{-5}$
    (solid blue line).  Vertical line at $T\approx 150\mev$ marks $\gamma =1$,
    $\dm(150\mev) \simeq 2.9\times 10^{-6}$. The red dashed line shows
    evolution of the chemical potential for $\Gamma_f = 0$. The green
    short-dashed line shows $\dm(\eta)$ for $r_B = 5\times 10^{-4}$. The
    purple dashed-dotted line shows the decay of the chemical potential in the
    absence of magnetic field.)}
  \label{fig:mu_f10_rg5_ns3}
\end{figure}

Finally, the exact numerical solution of the full system with continuous
spectrum and finite $\gamma(\eta)$ is shown in Fig.~\ref{fig:mu_f10_rg5_ns3}
where the red dashed line shows $\dm(\eta)$ in the case $\Gamma_f = 0$ and the
thick blue and green lines show $\dm(\eta)$ for $\Gamma_f \neq 0$ and
different $r_B$.  This full evolution follows that of $\Gamma_f = 0$ and then
breaks down exponentially fast when $\gamma(\eta) \sim 1$ and (\ref{eq:19})
holds.

\bigskip

\noindent \textbf{Conclusion.}  This work demonstrates that the
\emph{traditional MHD equations should be modified}, when applied to a plasma
of relativistic particles with $T\gg m$. The proper account of the
\emph{chiral anomaly} changes the evolution of magnetic fields in two ways:
\textit{(i)} the magnetic fields survive several orders of magnitude longer
than the time defined by magnetic diffusion
(Eqs.~(\ref{eq:17})--(\ref{eq:18})) and \textit{(ii)} an ``inverse cascade''
develops, transferring energy from shorter to longer wavelength modes. The
effect depends on the energy of magnetic fields parametrized by its ratio to
the total energy density, $r_B$.  In the literature discussing the evolution
of magnetic fields (see e.g.~\cite{Banerjee:03,Banerjee:04,Jedamzik:10}) $r_B
\le 1$ is often considered.  It was demonstrated e.g. in
Refs.~\cite{Baym:1995fk,Sigl:1996dm} that $r_B \sim \text{few}\times 10^{-3}$
may be generated at cosmological first order phase transitions.  The mechanism
of~\cite{Vachaspati:91} predicts maximally helical magnetic fields with $B
\sim 100\gev^2$ (i.e. $r_B \sim 10^{-2}$, c.f.~\cite{Chu:11}) at small scales.
See
\cite{Vachaspati:91,Baym:1995fk,Sigl:1996dm,Davidson:96,Copi:2008he,Kandus:10}
and refs. therein.

We refer to the value of $r_B$ at $T\sim 100\gev$.  Due to the inverse cascade
and helicity conservation this energy decreases by about an order of magnitude
by the time when our effect stops ($T\sim 10-100\mev$). The subsequent
evolution of the magnetic fields is described by the conventional
MHD~\cite{Campanelli:07,Banerjee:04,Giovannini:2003yn}, a significant part of
$r_B$ further dissipates, so that only the large scale tail of the spectrum
may survive due to turbulent effects. To predict the final fate of these CMF
for every initial spectrum and compare them with cosmological bounds (see
e.g.~\cite{Kahniashvili:10a}), our results should be combined with the MHD
analysis.  Nevertheless, the above-described mechanism, based \emph{entirely
  on the Standard Model}, clearly improves the chances of survival of CMF
generated at subhorizon scales~\cite{Durrer:2003ja}. Indeed, even for $r_B
\sim 10^{-5}$ the fields survive down to $T\sim 100\mev$, while for $r_B \sim
0.1$ the inverse cascade is operational down to $T\sim 10\mev$.  Moreover,
regardless of the survival of the CMF, this effect is important as the
left-right asymmetry in the electron sector survives down to
$T\sim\mathcal{O}(100)\mev$ and thus potentially affects important processes
in the early Universe: can change the nature of the QCD phase
transition~\cite{Schwarz:2009ii} and produce gravitational
waves~\cite{Durrer:2010xc}, leave its imprints on BBN and
CMB~\cite{Lesgourgues:99,Mangano:10}.


\bigskip

\textbf{Acknowledgments.} We would like to thank V.~Cheianov, B.~Pedrini, and
M.~Shaposhnikov for useful discussions.

\let\jnlstyle=\rm\def\jref#1{{\jnlstyle#1}}\def\aj{\jref{AJ}}
  \def\araa{\jref{ARA\&A}} \def\apj{\jref{ApJ}\ } \def\apjl{\jref{ApJ}\ }
  \def\apjs{\jref{ApJS}} \def\ao{\jref{Appl.~Opt.}} \def\apss{\jref{Ap\&SS}}
  \def\aap{\jref{A\&A}} \def\aapr{\jref{A\&A~Rev.}} \def\aaps{\jref{A\&AS}}
  \def\azh{\jref{AZh}} \def\baas{\jref{BAAS}} \def\jrasc{\jref{JRASC}}
  \def\memras{\jref{MmRAS}} \def\mnras{\jref{MNRAS}\ }
  \def\pra{\jref{Phys.~Rev.~A}\ } \def\prb{\jref{Phys.~Rev.~B}\ }
  \def\prc{\jref{Phys.~Rev.~C}\ } \def\prd{\jref{Phys.~Rev.~D}\ }
  \def\pre{\jref{Phys.~Rev.~E}} \def\prl{\jref{Phys.~Rev.~Lett.}}
  \def\pasp{\jref{PASP}} \def\pasj{\jref{PASJ}} \def\qjras{\jref{QJRAS}}
  \def\skytel{\jref{S\&T}} \def\solphys{\jref{Sol.~Phys.}}
  \def\sovast{\jref{Soviet~Ast.}} \def\ssr{\jref{Space~Sci.~Rev.}}
  \def\zap{\jref{ZAp}} \def\nat{\jref{Nature}\ } \def\iaucirc{\jref{IAU~Circ.}}
  \def\aplett{\jref{Astrophys.~Lett.}}
  \def\apspr{\jref{Astrophys.~Space~Phys.~Res.}}
  \def\bain{\jref{Bull.~Astron.~Inst.~Netherlands}}
  \def\fcp{\jref{Fund.~Cosmic~Phys.}} \def\gca{\jref{Geochim.~Cosmochim.~Acta}}
  \def\grl{\jref{Geophys.~Res.~Lett.}} \def\jcp{\jref{J.~Chem.~Phys.}}
  \def\jgr{\jref{J.~Geophys.~Res.}}
  \def\jqsrt{\jref{J.~Quant.~Spec.~Radiat.~Transf.}}
  \def\memsai{\jref{Mem.~Soc.~Astron.~Italiana}}
  \def\nphysa{\jref{Nucl.~Phys.~A}} \def\physrep{\jref{Phys.~Rep.}}
  \def\physscr{\jref{Phys.~Scr}} \def\planss{\jref{Planet.~Space~Sci.}}
  \def\procspie{\jref{Proc.~SPIE}} \let\astap=\aap \let\apjlett=\apjl
  \let\apjsupp=\apjs \let\applopt=\ao

\newpage \onecolumngrid
\appendix

\begin{center}
  \bfseries \large Supplementary material
\end{center}

\setcounter{figure}{0} %
\makeatletter %
\@addtoreset{figure}{section} \makeatother
\renewcommand{\thefigure}{\thesection\arabic{figure}}

\section{Evolution of a single mode}
\label{sec:evol-single-mode}

\begin{figure}[t]
  \centering
  \includegraphics[width=0.5 \textwidth]{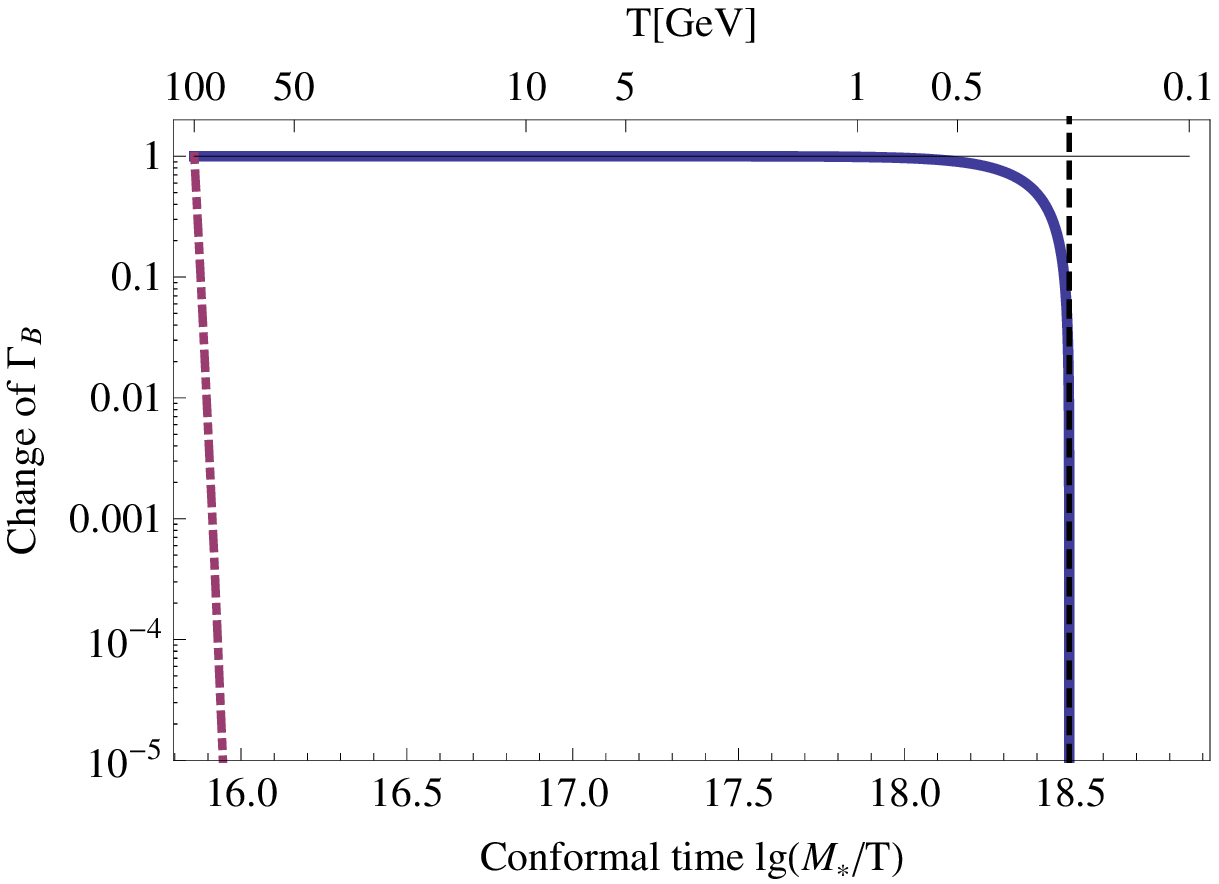}
  \caption{The relative change of helicity (solid line) as compared to the
    solution of~(\ref{eq:17}) (dashed line). The dot-dashed line shows the
    evolution of $\Gamma_B$ in the absence of chemical potential (solely due
    to magnetic diffusion with $\lg(\eta_\sigma) \approx 16.1$). A black
    dashed-dotted line shows the evolution of $\gamma(\eta)$, growing at
    $\eta^2$ at small times and then starting to increase exponentially fast.
    A thin black vertical line marks the solution of~(\ref{eq:19}).}
  \label{fig:helicity_flip_gamma1}
\end{figure}

This Appendix provides numerical results, illustrating analytic
results~(\ref{eq:15}), (\ref{eq:17}), and (\ref{eq:19}).
Fig.~\ref{fig:helicity_flip_gamma1} shows the relative change of $\Gamma_B$
from its initial value and the evolution of $\gamma(\eta)$.  We see that for
$\gamma\ll 1$ the value of $\Gamma_B$ (i.e. the total energy density of the
CMF) decays over a much longer time, $\eta_\sigma/\gamma$, than that of
magnetic diffusion, and, a significantly positive value of the chemical
potential (which would decay in the absence of magnetic field in a time
$\eta\sim \Gamma_f^{-1}$) is maintained during this time.

The evolution of the mode is therefore similar to the one discussed above, as
long as $\gamma \ll 1$. For $\gamma \sim 1$ the evolution quickly becomes that
of standard MHD without a difference of chemical potentials, with $\Gamma_B$
decaying due to magnetic diffusion and $\dm$ dissipating through perturbative
flipping.

\section{Evolution of several discrete modes ($\Gamma_f = 0$)}
\label{sec:energy-evolution-analytic-estimate}

\subsection{Analytic estimate of the time of the energy/helicity transfer
  between two modes}

The results for the time of transfer of the energy and helicity between two modes can be derived
in a different way than done in the main text. Namely, consider the initial spectrum
\begin{equation}
  \label{eq:56}
  H(\eta_0) = H_1^0\delta(k-k_1) + H_2^0 \delta(k-k_2)
\end{equation}
The spectrum preserves its shape throughout evolution and the tracking
solution of Eq.~(\ref{eq:20}) is given by
\begin{equation}
  \label{eq:57}
  \Delta\mu_\tr(\eta) \approx \frac{2\pi}\alpha\frac{k_1^2
    H_1 + k_2^2 H_2}{k_1   H_1 + k_2 H_2}
\end{equation}
In the presence of such a $\dm$, the two modes in~(\ref{eq:56}) evolve as
follows:
\begin{align}
  \label{eq:58}
  \dot H_1 &= -\frac{2k_1k_2}{\sigma_c} \frac{k_1-k_2}{k_1 H_1 + k_2 H_2}H_1
  H_2\quad,\quad \dot H_2 = -\dot H_1
\end{align}
To see how fast the helicity (energy) gets transferred from the mode $k_1$ to
$k_2$ let us combine two equations~(\ref{eq:58}) into a single equation for
the ratio $h(\eta) = H_1(\eta)/H_2(\eta)$ (such that $h(\eta_0)= h_i > 1$ and
$h(\eta)\to 0$ as $t\to \infty$):
\begin{equation}
  \label{eq:59}
  \frac{dh}{dt} = -\frac{2k_1^2}{\sigma}q(1-q)\frac{h(1+h)}{h+q}
\end{equation}
(where $q=k_2/k_1<1$). As a result we can obtain the time
difference for $h$ to change from $h_i$ to $h$:
\begin{equation}
  \label{eq:60}
  \Delta t = \frac{\sigma}{2k_1^2} \left[\frac1{1-q}\log\frac{h_i}{h}
    +\frac1q\log\frac{1+h_i}{1+h}\right]
\end{equation}
We can see that the time when helicities of both modes
equalize (i.e. $h(\eta)=1$) is given by
\begin{equation}
  \label{eq:61}
  \Delta t_H = \frac{\sigma}{2k_1^2} \left[\frac{\log h_i}{1-q}
    +\frac1q\log\frac{1+h_i}2\right]
\end{equation}
From Eq.(\ref{eq:60}) one can also easily determine the
time when the energies of two modes equalize (i.e. when
$h=\frac1q$).  

In the case when two modes with very different wave numbers
(i.e. when $q\ll 1$), the time $\Delta t_H$ is approximately
equal to
\begin{equation}
  \label{eq:62}
  \Delta t_H \approx \frac{\sigma}{2k_1^2}\frac1q\log
  h_i = \frac{2\pi\sigma}{k_1 k_2}\log h_i
\end{equation}
Notice, that this time is much longer than the magnetic diffusion time
$\frac{\sigma}{2k_1^2}$ and it depends on the ratio of initial amplitudes only
logarithmically.

\subsection{Approximations, used in derivation of Eq.~(\protect\ref{eq:21})}
\label{sec:appr-used-paper}

In this Section we provide some details regarding the derivation, used in the
main paper and leading to Eq.~(\ref{eq:21}). Consider the system of
Eqs.~(\ref{eq:20}). Let us assume that $k_1 H_1(\eta_0) \gg k_2 H_2(\eta_0)$
and $k_2 < k_1$. Then, we can rewrite the tracking solution~(\ref{eq:57}) as
\begin{equation}
  \label{eq:50}
  \dm_\tr \approx \frac{2\pi k_1}{\alpha}\left(1 -\epsilon\right)
\end{equation}
(where $\epsilon = \frac{k_2 H_2}{k_1 H_1} \ll 1$). Its time evolution is
given by
\begin{equation}
  \label{eq:51}
  \frac{d\dm}{dt} = -\Gamma_B\left(\dm - \frac{2\pi k_1}{\alpha}(1-\epsilon(\eta))\right)
\end{equation}
The evolution of two modes $H_1$ and $H_2$ is then given by
\begin{equation}
  \label{eq:52}
  \dot H_1 = -\frac{\epsilon(\eta)}{t_{\sigma1}} H_1 = -\frac{k_2}{k_1
    t_{\sigma1}} H_2
\end{equation}
and
\begin{equation}
  \label{eq:53}
  \dot H_2 = \frac{H_2}{t_{\sigma2}} \left(\frac{k_1\bigl(1-\epsilon(\eta)\bigr)}{k_2}-1\right)
\end{equation}
from which it follows that
\begin{equation}
\label{eq:54}
H_2(\eta) = H_2(\eta_0)e^{\frac{k_1}{k_2}\frac{\eta}{\eta_{\sigma2}}}
\end{equation}
This result is confirmed by numerical solution in
Fig.~\ref{fig:nof2_vs_linear}, where the solution of Eq.~(\ref{eq:54}) is
shown in black dot-dashed line (see also~\ref{fig:nof10_vs_nof2}).  From here
we can find the equation for $\dot\Gamma_B$:
\begin{equation}
  \label{eq:55}
  \dot \Gamma_B = \left(\frac{2c_\Delta\alpha^2}{\pi
      \sigma_c}\right)\frac{k_1-k_2}{\sigma_c} k_1k_2 H_2
\end{equation}

\subsection{Evolution of several discrete modes}
\label{sec:nof10_nof2}

\begin{figure*}[t]
  \centering
  \begin{tabular}{cc}
    \includegraphics[width=.5\textwidth]{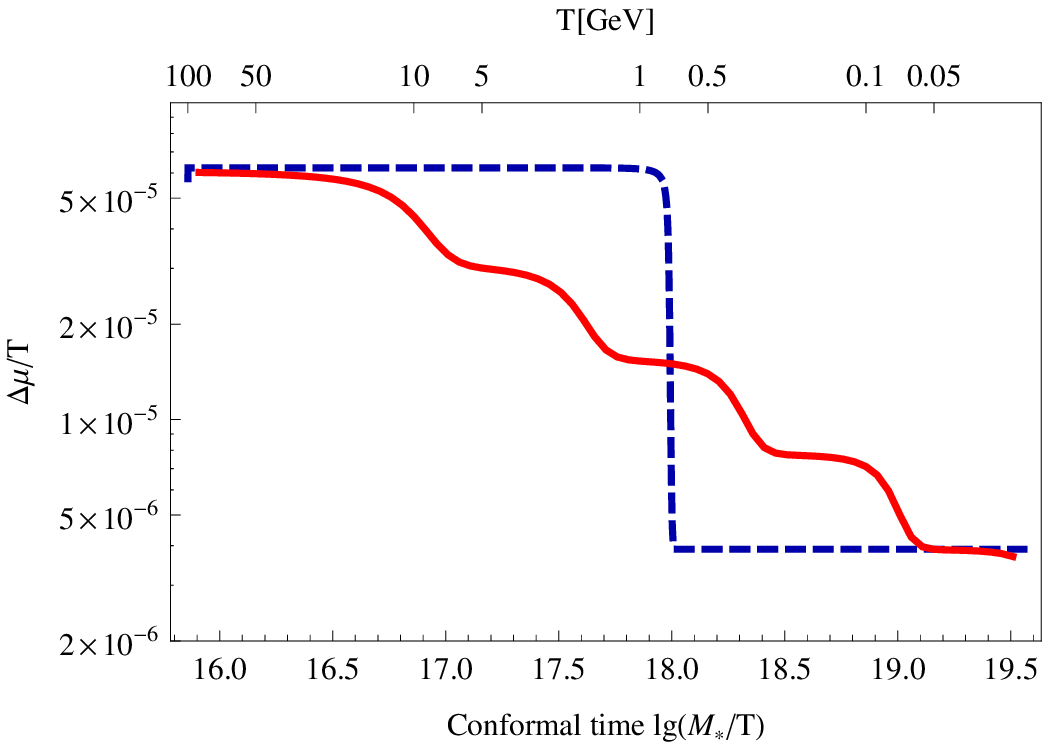} &
    \includegraphics[width=.5\textwidth]{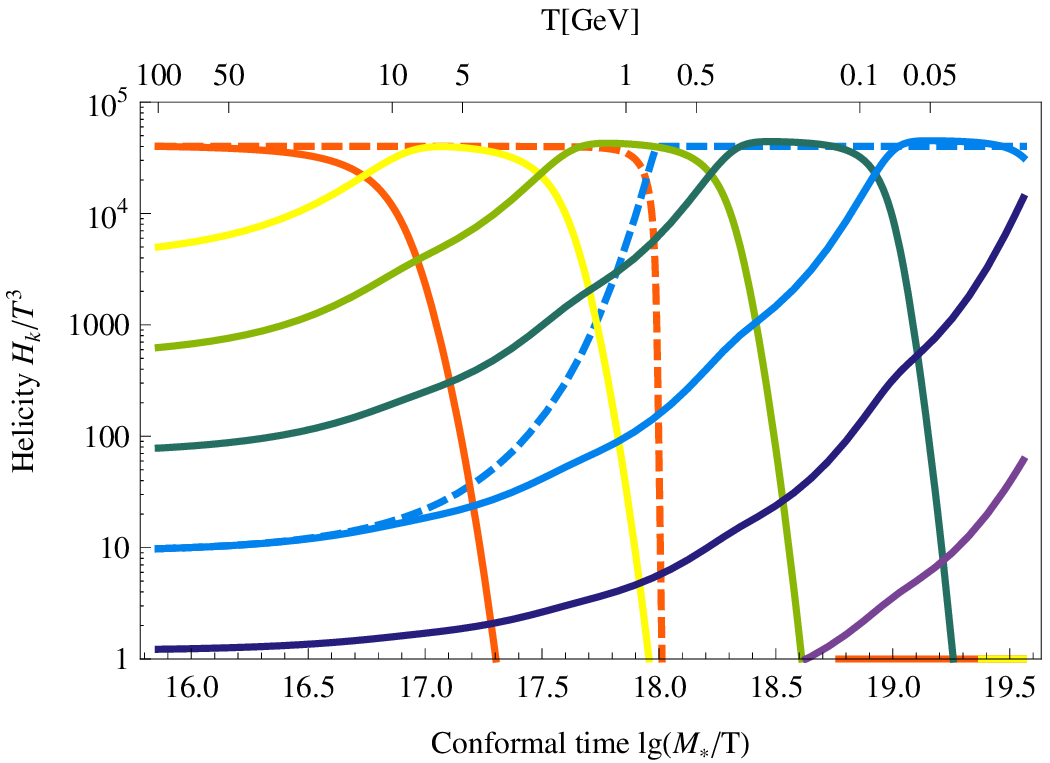}
  \end{tabular}
  \caption{Comparison between evolution of chemical potential (\textbf{left})
    and helicity modes (\textbf{right}) for 10 modes (solid lines) and 2 modes
    (1st and 5th of the previous set) -- dashed lines.}
  \label{fig:nof10_vs_nof2}
\end{figure*}

\begin{figure}
  \centering
  \includegraphics[width=.5\textwidth]{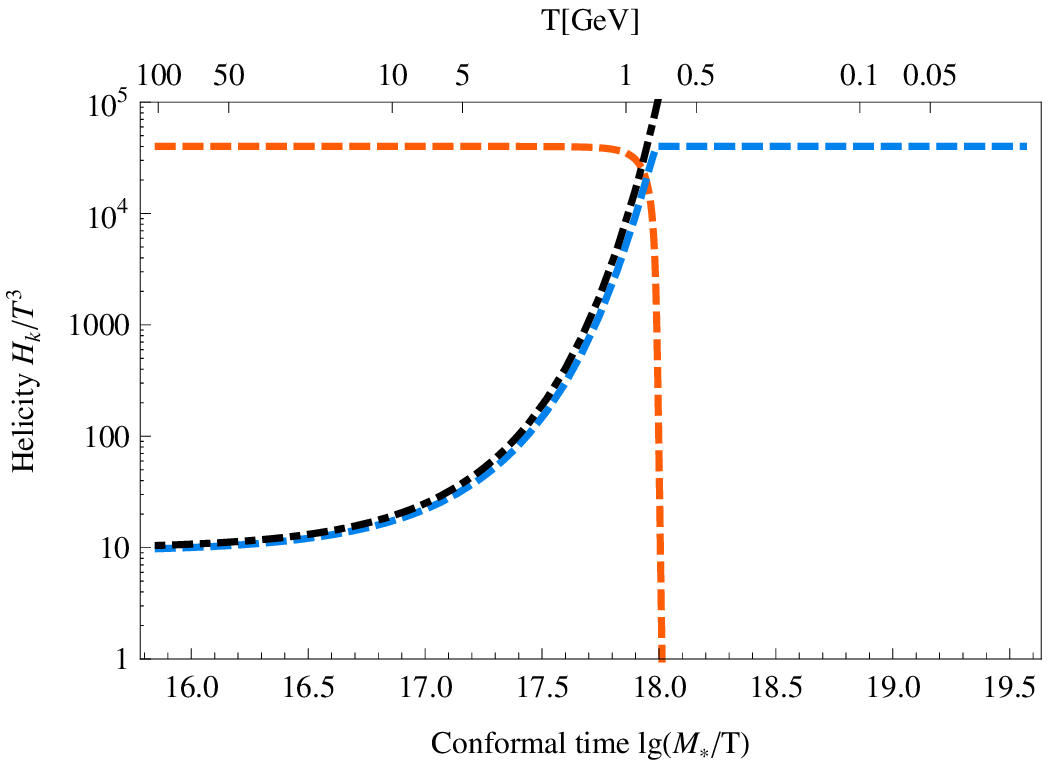}
  \caption{Comparison of the evolution of a long-wavelength mode (in the 2
    mode solution shown in Fig.\protect\ref{fig:nof10_vs_nof2}) with its
    linear approximation, given by Eq.~(\ref{eq:21}) -- dashed-dotted black
    line.}
  \label{fig:nof2_vs_linear}
\end{figure}

The evolution of 10 modes, sampling a continuous spectrum with $\CH_k^0
\propto k^3$ without perturbative chirality flip (solid lines in
Fig.~\ref{fig:nof10_vs_nof2}) and its comparison with the evolution of the
same spectrum but for two modes only with the same helicities initially stored
in each mode -- short-dashed lines in
Fig.~\ref{fig:nof10_vs_nof2}. Fig.~\ref{fig:nof2_vs_linear} compares the exact
numerical solution of the case with two modes only (red and blue lines) with
the result~(\ref{eq:21}) of the approximate (in $\CH_{k_2}/\CH_{k_1}$)
approximation (black dot-dashed line).

\section{Inverse cascade and the shape of the resulting spectrum}
\label{sec:shape-total-spectrum}

\begin{figure*}[t]
  \centering
  \begin{tabular}{cc}
  \includegraphics[width=.5\textwidth]{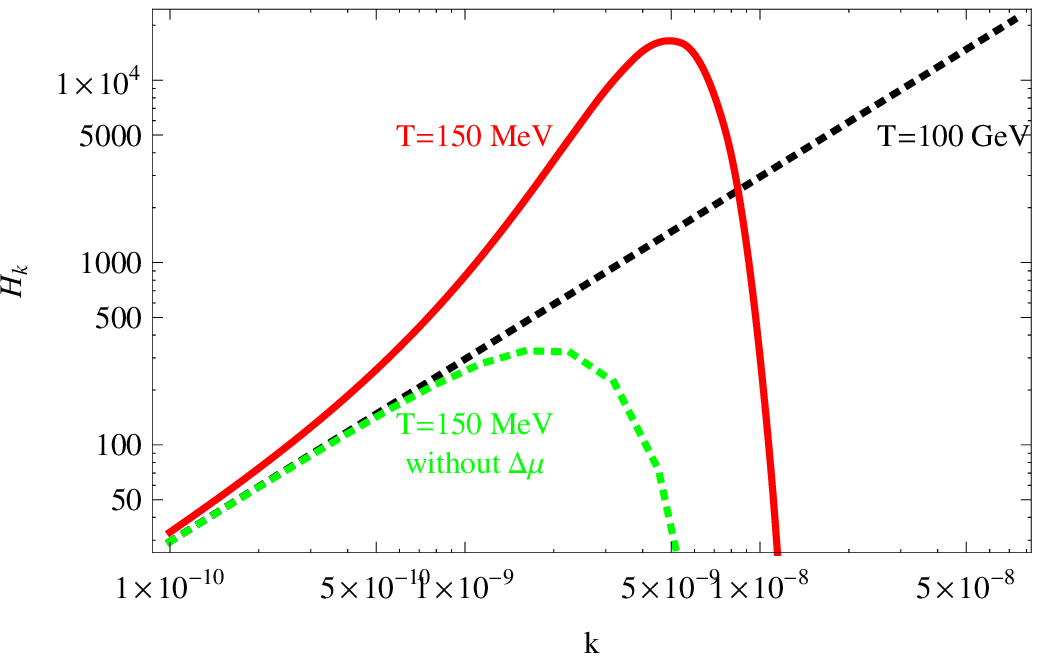} &
  \includegraphics[width=.5\textwidth]{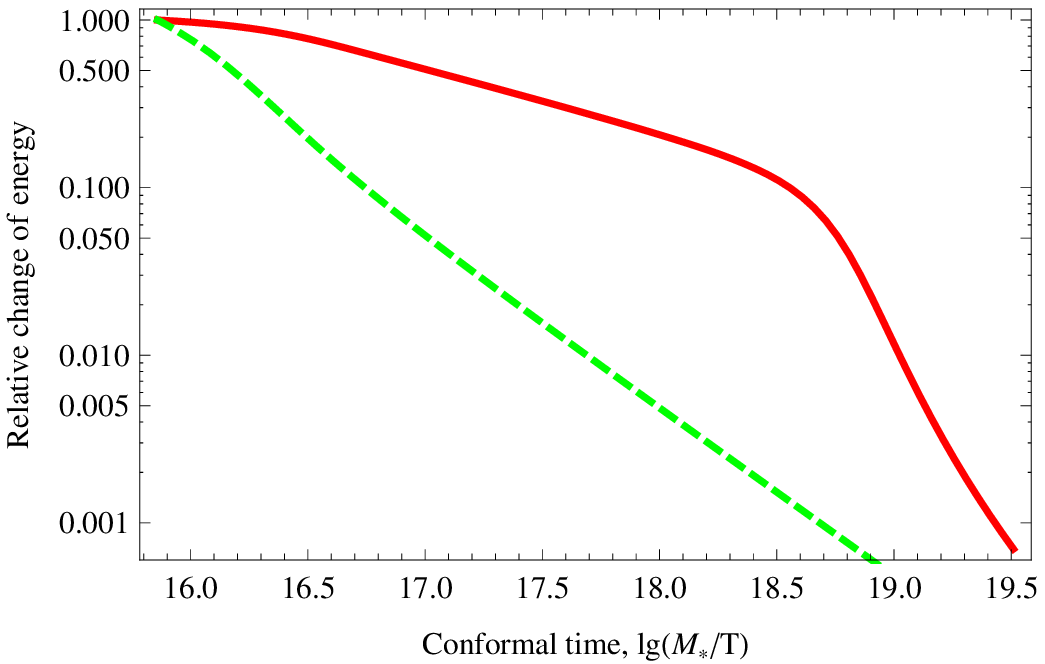}  
  \end{tabular}
  \caption{\textbf{Left:} Initial helicity spectrum $\CH_k^0 \propto k$ (black
    dotted line), the evolved spectrum $\CH_k$ at $T\approx 150\mev$ when
    $\gamma = 1$ (red solid line) as well as the helicity spectrum evolved
    \emph{solely} due to the magnetic diffusion (green dashed line) for $r_B
    \approx 5\times 10^{-5}$. \textbf{Right:} The relative change of the total
    energy with (red, solid) and without (green, dashed) the chemical
    potential difference.}
  \label{fig:Hk_f10_rg5_ns3}
\end{figure*}

In the case of the continuous spectrum while $\gamma \ll 1$ the evolution is
identical to the case of $\Gamma_f = 0$, described above (see
Fig.~\ref{fig:mu_f10_rg5_ns3}).  The condition $\gamma(\eta) \le 1$ determines
the time over which the described effect is present in plasma. We see from
Eq.~(\ref{eq:18}) that for $r_B \gtrsim 10^{-5}$ the fields, generated very
early (probably at $T_0 \sim 100\gev$) can survive until $T\sim 100\mev$ and
during the same time the non-zero chemical potential remains in plasma.  The
value of $k$ at which the evolution stops can be determined via~(\ref{eq:19})
with $\gamma(\eta) \sim 1$.  We find that $k\approx k_\max(\eta)$ -- the
wave-number at which the spectrum is peaked at the moment $\eta$ is of the
order of final time, determined by~(\ref{eq:18}), i.e.
$\eta_\sigma(k_\max(\eta)) \sim \eta$. Thus the value of $k_\max$ is
approximately the same as in the standard MHD case, when all modes with
$\eta_\sigma(k) < t$ would be erased by the magnetic
diffusion. \emph{However}, the total energy of the spectrum (or, equivalently
$\CH_{k_\max}$ is much higher in our case -- the presence of chemical
potential allows for an ``inverse cascade'' process -- the energy stored in
short wave-lengths modes to be transferred to the longer wave-lengths (as
Fig.~\ref{fig:Hk_f10_rg5_ns3}).

\section{Perturbative chirality flipping rates}
\label{sec:pert-chir-flipp}

\begin{figure}[t]
  \centering
  \includegraphics[width=.5\textwidth]{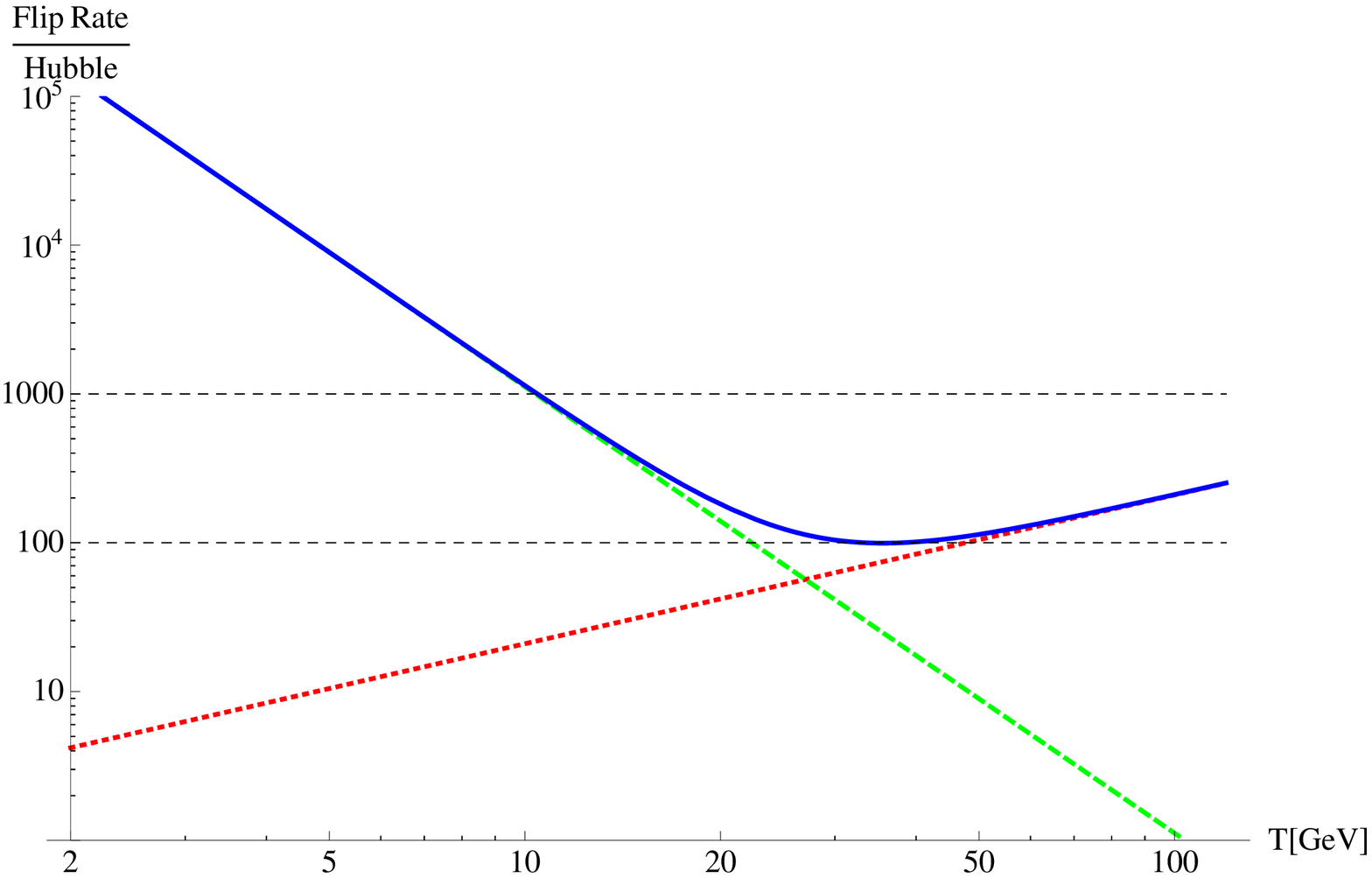}
  \caption{Ratio of chirality-flipping rates to the Hubble rate for $T$ in GeV
    range (blue solid line). Red dashed line is the flipping rate due to weak
    reactions, and green (dotted) is the rate due to electromagnetic
    processes.}
  \label{fig:flip-rates}
\end{figure}

Fig.~\ref{fig:flip-rates} shows the ratio of perturbative chirality-flipping
rates due to weak or electromagnetic reactions to the Hubble expansion rate
(as a function of time).

\section{Derivation of equations for $\CH_k$}
\label{sec:derivation-eq-Hk}

In this Appendix we provide the details of derivation of
Eqs.~(\ref{eq:6})--(\ref{eq:7}).

Due to 3d translation and rotation invariance, we define the Fourier modes
$B_k$ and $A_k$ for the magnetic field and its gauge potential in the usual
way
\begin{equation}
  \label{eq:22}
  B(x) = \int \frac{d^3k}{(2\pi)^3}\, e^{i k\cdot x} B_k
\end{equation}
and introduce \emph{magnetic helicity density} $\CH_k$ and \emph{magnetic
  energy density} $\rho_k$ in the $k$-space:
\begin{equation}
  \label{eq:23}
  \CH(\eta) = \int \frac{d^3k}{(2\pi)^3}\, \vec A_k \cdot \vec B_k^* \equiv \int dk\,\CH_k 
  \end{equation}
\begin{equation}
  \label{eq:24}
 \rho_B \equiv \frac1{2V}\int \frac{d^3k}{(2\pi)^3}\, \bigl|\vec B_k\bigr|^2 = \int dk\, \rho_k
\end{equation}
( $\vec B_k^* = \vec B_{-k}$ as $B$ is real). Notice that definitions of
$\CH_k$ and $\rho_k$ in Eqs.~(\ref{eq:23})--(\ref{eq:24}) contain integrals
over the absolute value of the 3-vector $k$ only
(cf.~\cite{Banerjee:03,Banerjee:04,Campanelli:07}.  Multiplying Fourier
version of Eq.~(\ref{eq:5}) by the complex-conjugated mode $\vec B_k^*$, we
get:
\begin{equation}
  \label{eq:25}
  \dot{\vec B}_k\vec  B_k^* + \dot {\vec B}_k^* \vec B_k =
  -\frac{2k^2}{\sigma_c}\vec B_k^*  \vec B_k + i
  \frac{2\alpha}{\pi}\frac{\Delta\mu}{\sigma_c} \vec B_k^*\cdot (\vec k \times \vec B_k)
\end{equation}
 With the use of $\vec B_k = i\vec k \times \vec A_k$ we obtain from
Eq.~(\ref{eq:25}):
\begin{align}
  \label{eq:26}
  \pfrac{\rho_k}{\eta} &= -\frac{2k^2}{\sigma_c}\rho_k +
  \frac{\alpha}{2\pi}\frac{\Delta\mu}{\sigma_c} k^2\CH_k \\
  \pfrac{\CH_k}{\eta} &= -\frac{2k^2}{\sigma_c}\CH_k +
  \frac{2\alpha}{\pi}\frac{\Delta\mu}{\sigma_c} \rho_k
\end{align}
If fields are maximally helical, i.e. $\rho_k = \frac{k}2\CH_k$,
Eq.~(\ref{eq:26}) reduces to~(\ref{eq:6}).

\section{Dependence on the parameters of the initial spectrum}
\label{sec:depend-param-init}

\begin{figure*}[t]
  \centering
  \includegraphics[width=0.5\textwidth]{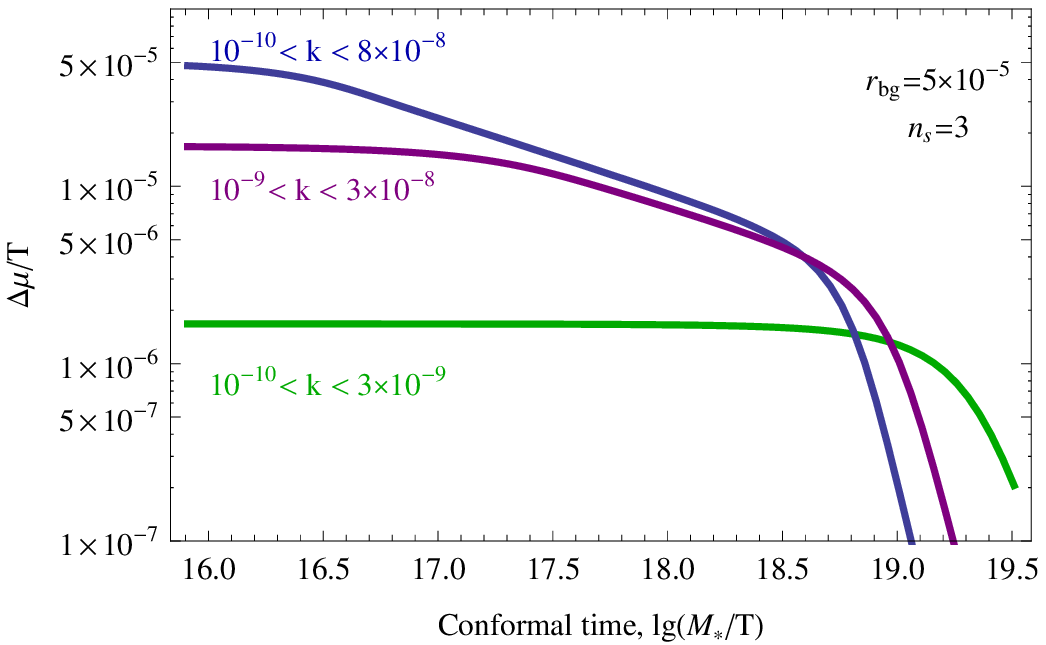}~\includegraphics[width=0.5\textwidth]{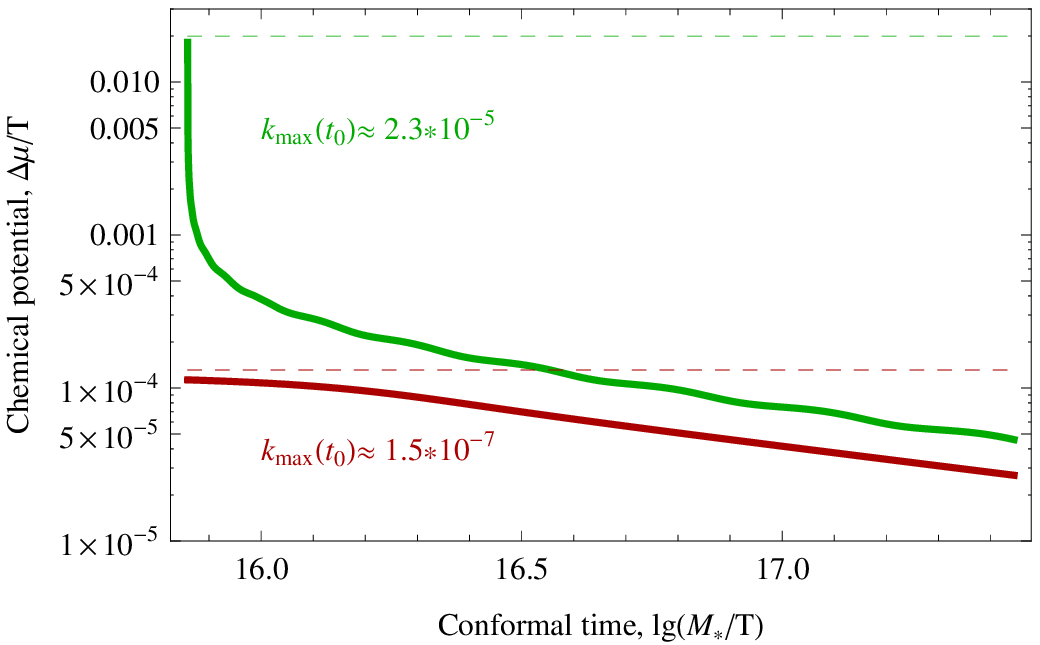}
  \label{fig:mu_rg5_ns3_different_kmax}
  \caption{\textbf{Left:} Evolution of $\dm$ for two spectra with the same
    total energy density $r_B \approx 5\times 10^{-5}$, the same initial
    spectral index $n_s=3$, and \textbf{different range of initial $k$}. $\lg
    \eta_\sigma=18.7$ (green), $\lg \eta_\sigma = 16.7$
    (purple). \textbf{Right:} Zoom on initial region of evolution of $\dm$ for
    two spectra with the same total energy density $r_B \approx 1$ and the
    same initial spectral index ($n_s =5$) but \textbf{different range of
      initial $k$ over which this energy is distributed}.}
\end{figure*}

Figs.~\ref{fig:mu_rg5_ns3_different_kmax}1 show the dependence of the chemical
potential difference on the range of modes $k$ in the spectrum.

\section{Notations}

\subsection{Magnetic energy density}
\label{sec:energy-density}

In this Appendix we discuss several conventions of expressing the energy
density of the magnetic field $\rho_B = \frac 1{2V}\int d^3x\, B^2 $, used in
the literature.

One possibility (used in this paper) is to express it in terms of the total radiation energy
density
\begin{equation}
  \bar\rho \equiv \frac{\pi^2}{30}g_\eff
  T^4\label{eq:13}
\end{equation}
Alternatively, one can express $\rho_B$ in terms of the total entropy of
radiation $\bar s = \frac{2\pi^2}{45} g_\eff T^3$ as follows
(cf.~\cite{Banerjee:04}):
\begin{equation}
  \label{eq:29}
  r_g \equiv \frac{\rho_B}{\bar s^{4/3}}
\end{equation}
To convert between the $r_B$ and $r_g$ one can use:
\begin{equation}
  \label{eq:91}
  r_g \approx \frac{r_B}{g_\eff^{1/3}}\approx 0.2 r_B \parfrac{100}{g_\eff}^{1/3}
\end{equation}

\subsection{Dimensionless quantities}
\label{sec:dimens-notat}

Eqs.~(\ref{eq:4}--\ref{eq:5}) do not change in the expanding universe with FRW
metric: $ds^2 = -dt^2 + a^2(t)d\vec x^2 = a^2(\eta)(-d\eta^2 + d\vec x^2)$,
provided that we work with conformal coordinates $x$, use (dimensionless)
conformal time $\eta = \frac{M_*}T$ (where we have used $a(t) = 1/T$ and
neglected that $g_\eff$ changes with time) and substitute $\Delta\mu,\sigma,T,
\Gamma_f$ with their conformal counterparts $(\Delta \mu a), (\sigma a), (T
a), (\Gamma_f a)$, and introduce conformal electro-magnetic fields $E\to a^2 E
= E_c$, $B\to a^2 B = B_c$.  Notice that $\sigma a = \sigma_c \approx\const
$~\cite{Baym:1997gq}. We use these coordinates throughout the paper, starting
from Eq.~(\ref{eq:8}). To restore the dimensions it is sufficient to change
any quantity as follows: $\dm \to \dm/T$, $k\to k/T$, $\CH_k\to \CH_k/T^3$,
$\rho_k \to \rho_k/T^4$, etc.

\end{document}